\documentclass[aps,pra,twocolumn,groupedaddress,nofootinbib]{revtex4-1}
\usepackage{amsmath}
\usepackage{amssymb}
\usepackage{graphicx,subfigure}
\usepackage{mathrsfs}
\usepackage{ntheorem}
\usepackage[T1]{fontenc}
\usepackage{bm}
\usepackage{subfigure}
\usepackage[bookmarks=false]{hyperref}
\hypersetup{colorlinks=true,citecolor=blue,
linkcolor=red,urlcolor=blue,pdfstartview=FitH,
bookmarksopen=true}

\newcommand{\ket}[1]{|#1\rangle}
\newcommand{\bra}[1]{\langle #1|}
\newcommand{\inp}[2]{\langle #1|#2\rangle}
\newcommand{\inpp}[2]{\prec #1,#2 \succ}

\newcommand{\tr}{\mathrm{tr}}
\newcommand{\ud}{\mathrm{d}}

\newcommand{\abs}[1]{\lvert #1\rvert}

\def\CC{{\rm\kern.24em \vrule width.04em height1.46ex depth-.07ex \kern-.30em C}}
\def\RR{{\rm\kern.24em \vrule width.04em height1.46ex depth-.07ex
\kern-.30em R}}
\def\P{{\rm I\kern-.25em P}}

\begin{document}
\title{Differential geometry of time-dependent $\mathcal{PT}$-symmetric quantum mechanics}

\author{Da-Jian Zhang}
\affiliation{Department of Physics, National University of Singapore, Singapore 117542}
\author{Qing-hai Wang}
\affiliation{Department of Physics, National University of Singapore, Singapore 117542}
\author{Jiangbin Gong}
\email{phygj@nus.edu.sg}
\affiliation{Department of Physics, National University of Singapore, Singapore 117542}
\date{\today}

\begin{abstract}
Time-dependent $\mathcal{PT}$-symmetric quantum mechanics is featured by a varying inner-product metric and has stimulated a number of interesting studies beyond conventional quantum mechanics. In this paper, we explore geometric aspects of time-dependent $\mathcal{PT}$-symmetric quantum mechanics. We not only find a geometric phase factor emerging naturally from cyclic evolutions of $\mathcal{PT}$-symmetric systems, but also formulate a series of differential geometry concepts, including  connection, curvature, parallel transport, metric tensor, and quantum geometric tensor. Our findings constitute a useful, perhaps indispensible, tool to
tackle physical problems involving $\mathcal{PT}$-symmetric systems with time-varying system's parameters. To exemplify the application of our findings, we show that the unconventional geometrical phase [Phys. Rev. Lett. \textbf{91}, 187902 (2003)], consisting of a geometric phase and a dynamical phase proportional to the geometric phase, can be expressed as a single geometric phase identified in this work.
\end{abstract}

\maketitle

\section{Introduction}

Standard quantum mechanics for bounded states is built upon a fixed Hilbert space, with the associated inner product of two complex vectors defined by the Dirac bra-ket notation.  However, such a quantum mechanics may not consistently treat physical problems with varying Hilbert spaces. For example, it is not obvious how to depict the dynamics of a particle in an infinitely deep square-well potential with a moving boundary, of which the instantaneous Hilbert space changes with time.  It is thus necessary and motivating to formulate a new type of quantum mechanics that allows the inner product structure to change along a parameter path.  In particular, following the pioneering work of Bender and Boettcher concerning time-independent parity-time-reversal-symmetric ($\mathcal{PT}$-symmetric) quantum mechanics ($\mathcal{PT}$QM) \cite{1998Bender5243}, Gong and Wang put forward time-dependent $\mathcal{PT}$QM \cite{2013Gong485302},  featured by a Schr\"{o}dinger-like equation explicitly accounting for  a varying inner-product metric. Time-dependent $\mathcal{PT}$QM has spurred reexaminations of interesting issues in statistics mechanics \cite{2018Wei12105,2018Wei12114,2015Deffner150601,2017Zeng31001}
and quantum dynamics \cite{2016Fring42128, 2016Fring42114,2017Maamache383,2018Mostafazadeh46022,2016Gardas23408,2017Mead85001}, such as the Jarzynski equality \cite{2015Deffner150601,2017Zeng31001}, the Carnot bound \cite{2016Gardas23408}, and the selection rule \cite{2017Mead85001}.

In this work as an accompanying paper to
Ref.~\cite{2018Zhang}, we focus on the topic of geometric aspects of time-dependent $\mathcal{PT}$QM. This topic is a fascinating subject.  First, given the fact that a varying metric is always excluded in standard quantum mechanics, its intriguing interplay with other concepts in quantum physics, e.g., the Berry phase \cite{1984Berry45}, is still in its infancy. Second, even in the absence of a varying metric, geometric aspects of standard quantum mechanics are known to be of profound importance in various frontier topics of quantum computation, quantum information, and condensed-matter physics  \cite{1989Shapere, 2003Bohm}. One thus anticipates that physics arising from a varying metric shall advance our fundamental understanding of the profound role of geometry in time-dependent $\mathcal{PT}$QM. Last but not least, in view of ongoing investigations of physical properties, especially topological properties, of $\mathcal{PT}$-symmetric systems \cite{2017Xiao1117,
2017Ashida15791,2017Kawabata190401,
2017Weimann433,2017Menke174506a,
2018Kawabata85116,2018Lourenco85126,2018Shen146402,
2018Yao136802,2018Gong31079,2018El-Ganainy11}, a systematic inspection of geometric aspects of time-dependent $\mathcal{PT}$QM would be a useful, perhaps indispensible, reference point to
tackle physical problems involving time-varying system's parameters, which are frequently encountered in the current research of $\mathcal{PT}$-symmetric systems.

For the above reasons, the purpose of this work is to present comprehensive and rigorous results regarding geometric aspects of time-dependent $\mathcal{PT}$QM. To this end, we start with the identification of a geometric phase (GP) that emerges naturally from a cyclic evolution of a $\mathcal{PT}$-symmetric system. Then, with the motivation of revealing the geometry underlying our GP, we formulate, in succession, a series of differential geometry concepts, including  connection, curvature, parallel transport, metric tensor, and quantum geometric tensor (QGT). Almost all of these concepts bear some resemblance to their respective counterparts in standard quantum mechanics.
In particular, the QGT advocated here, similar to its counterpart \cite{1980Provost289}, is a complex Hermitian tensor, with its imaginary part giving our curvature and its real part inducing our metric tensor on system's parameter manifold. On the other hand, our metric tensor, however, may be Riemannian or pseudo-Riemannian, depending on the physical context under consideration. Its pseudo-Riemannian feature is absent in standard quantum mechanics.
To exemplify the application of our findings, we revisit one well-known example displaying the so-called unconventional GP \cite{2003Zhu187902},
which consists of a GP and a dynamical phase (DP) proportional to the GP.  We show that the unconventional GP, instead of being the sum of a GP and a DP, can be expressed as a single GP found in this paper, with the associated metric tensor categorized as a pseudo-Riemannian metric elusive in standard quantum mechanics.

Results of this paper can be regarded as a more general
reformulation as well as an extension of Ref.~\cite{2018Zhang}. Unlike those in Ref. \cite{2018Zhang}, which depict geometric aspects of the eigenstates of a $\mathcal{PT}$-symmetric Hamiltonian, the results of this paper are applicable to a more general physical context, e.g., the involved quantum states may not be eigenstates of the Hamiltonian.

This paper is organized as follows. In Sec. \ref{sec:fundamentals}, we recapitulate some fundamentals of $\mathcal{PT}$QM. In Sec. \ref{sec:GP}, we identify the GP. In Sec. \ref{sec:concepts}, we formulate a series of differential geometry concepts, including  connection, curvature, parallel transport, metric tensor, and QGT. Section \ref{sec:unconventional GP} presents our interpretation  of the unconventional GP, and Sec. \ref{sec:conclusion} concludes this work with some necessary remarks.

\section{From time-independent $\mathcal{PT}$QM to time-dependent $\mathcal{PT}$QM}\label{sec:fundamentals}

Consider a quantum system with a Hilbert space $\mathcal{H}$. For simplicity, we assume that $\textrm{dim}(\mathcal{H})<\infty$, but actually, our discussion may be extended to the case of infinite dimension, as can be seen in the example in Sec. \ref{sec:unconventional GP}. In order to ensure
the applicability of standard quantum measurement theory, it is necessary for the Hamiltonian, denoted by $H$, of the system to be diagonalizable and with a real spectrum \cite{2006Mostafazadeh919}. Since the work of  Bender and Boettcher \cite{1998Bender5243}, it has been realized that such a necessary condition can be satisfied even if $H$ is not Hermitian. Indeed, $H$ is diagonalizable and with a real spectrum if and only if there exists a positive-definite operator $W$ such that $WH=H^\dagger W$ \cite{2002Mostafazadeh3944}. If such an operator is found, a consistent quantum theory, i.e., time-independent $\mathcal{PT}$QM, can be built. In this theory, the physical Hilbert space is obtained by endowing $\mathcal{H}$ with a new inner product
$\inpp{\cdot}{\cdot}:=\bra{\cdot}W\ket{\cdot}$. Accordingly, a Hermitian operator $X$ over the physical Hilbert space, referred to as physical Hermitian operator for convenience, satisfies $\inpp{\cdot}{X\cdot}=\inpp{X\cdot}
{\cdot}$ or equivalently $WX=X^\dagger W$. In this language, $H$ is a physical Hermitian operator, and moreover, any observable in this theory is identified with some physical Hermitian operator thus defined.

Gong and Wang \cite{2010Gong12103,2013Gong485302} considered the scenario where $H$ depends on some system's parameters, denoted collectively as $\lambda$, i.e., $H=H(\lambda)$.
Here, the system's parameters belong to a manifold $M$, which may arise from the classical configuration of control fields. Accordingly, $\lambda=(\lambda^1,\cdots,\lambda^m)$ with $m=\textrm{dim}(M)$. As in time-independent $\mathcal{PT}$QM, $H(\lambda)$ is required to fulfill the condition that $W(\lambda)H(\lambda)=H^\dagger(\lambda) W(\lambda)$ for a positive-definite operator $W(\lambda)$, which depends on $\lambda$, too. Then, the physical Hilbert space, denoted as $\mathcal{H}(\lambda)$, is endowed with the inner product
$\inpp{\cdot}{\cdot}_\lambda:=\bra{\cdot}W(\lambda)\ket{\cdot}$,
referred to as the $\lambda$-dependent inner product hereafter. This scenario
is also the focus of this paper.

Gong and Wang then considered evolution problems where $\lambda\in M$ changes with time, i.e., $H(t)=H(\lambda_t)$, with $\lambda_t\in M$ varying over a time interval $[0,\tau]$. For this, the physical Hilbert space moves with time, and the evolving state $\ket{\psi(t)}$ at time $t$
belongs to $\mathcal{H}(\lambda_t)$. The Schr\"{o}dinger-like equation yielding unitary evolution is found to be ($\hbar=1$)
\begin{eqnarray}\label{eq:Sch}
i\partial_t\ket{\psi(t)}=\left[H(t)+iK(t)\right]\ket{\psi(t)},
\end{eqnarray}
where
\begin{eqnarray}\label{eq:gauge-field}
K(t)=-\frac{1}{2}W^{-1}(\lambda_t)\partial_tW(\lambda_t)
\end{eqnarray}
is a physical Hermitian operator, representing a gauge field necessary for unitarity. That is, the $\lambda$-dependent inner product of two arbitrary initial states is preserved during the evolution.
The expression (\ref{eq:gauge-field}) of $K(t)$ has been also justified by others \cite{2018Mostafazadeh46022}.

\section{Geometric phase}\label{sec:GP}

In the following, $\mathcal{N}(\lambda)$ denotes the set of normalized states in $\mathcal{H}(\lambda)$, i.e., $\mathcal{N}(\lambda):=\{\ket{\psi}\in\mathcal{H}(\lambda) |\inpp{\psi}{\psi}_\lambda=1\}$. Physically speaking, two states $\ket{\psi}\in\mathcal{N}(\lambda)$ and $\ket{\phi}\in\mathcal{N}(\lambda^\prime)$ are not comparable when $\lambda\neq\lambda^\prime$, since they belong to different physical Hilbert spaces. Therefore, $\ket{\psi}\in\mathcal{N}(\lambda)$ and $\ket{\phi}\in\mathcal{N}(\lambda^\prime)$ can be regarded as being identical if and only if $\lambda=\lambda^\prime$ and $\ket{\psi}=\ket{\phi}$.
For a state $\ket{\psi}\in\mathcal{N}(\lambda)$,
we introduce an associated state
$\ket{\widetilde{\psi}}:=W(\lambda)\ket{\psi}$. Here, the tilde is used to distinguish the state $\ket{\psi}$ from its associated state $\ket{\widetilde{\psi}}$. Accordingly, there is $\bra{\widetilde{\psi}}=\bra{\psi}W(\lambda)$. Using the associated state $\ket{\widetilde{\psi}}$, we can define an associated operator
$\rho:=\ket{\psi}\bra{\widetilde{\psi}}$.
It is not difficult to see that $\rho$ is a positive operator over $\mathcal{H}(\lambda)$ satisfying $\tr(\rho)=1$ and $\rho^2=\rho$, that is, it fulfills the conditions of being a density operator for a pure state. So, $\rho$ can be seen as the density operator associated to $\ket{\psi}\in\mathcal{N}(\lambda)$.
Similar to the two states $\ket{\psi}\in\mathcal{N}(\lambda)$ and $\ket{\phi}\in\mathcal{N}(\lambda^\prime)$, their associated density operators
$\rho=\ket{\psi}\bra{\widetilde{\psi}}$
and $\sigma=\ket{\phi}\bra{\widetilde{\phi}}$ can be regarded as being identical if and only if $\lambda=\lambda^\prime$ and $\ket{\psi}=e^{i\vartheta}\ket{\phi}$ for some $\vartheta\in\mathbb{R}$.

To arrive at our GP, suppose that the evolving state $\ket{\psi(t)}$ of the system returns to its initial physical state, i.e., $\ket{\psi(\tau)}=e^{i\alpha}\ket{\psi(0)}$, and moreover, the system's parameters return to their initial values, i.e., $\lambda_\tau=\lambda_0$. This defines a curve of density operators
\begin{eqnarray}\label{eq:curve}
C:t\in[0,\tau]\mapsto \rho(t),
\end{eqnarray}
with
\begin{eqnarray}\label{rhot}
\rho(t):=\ket{\psi(t)}\bra
{\widetilde{\psi}(t)}.
\end{eqnarray}
Since $\rho(0)$ and $\rho(\tau)$ are identical, $C$ in Eq. (\ref{eq:curve}) with Eq. (\ref{rhot}) represents a closed curve. Now, define an auxiliary state
$\ket{\phi_a(t)}:=e^{-if(t)}\ket{\psi(t)}$, with $f(\tau)-f(0)=\alpha$.
By definition, $\ket{\phi_a(\tau)}=\ket{\phi_a(0)}$. Substituting $\ket{\psi(t)}=e^{if(t)}\ket{\phi_a(t)}$ into Eq.~(\ref{eq:Sch}) and contracting its both sides with $\bra{\widetilde{\phi}_a(t)}$, we have
$\dot{f}(t)=-\inpp{\phi_a(t)}{[H(t)+iK(t)]
\phi_a(t)}_{\lambda_t}
+i\inpp{\phi_a(t)}{\dot{\phi}_a(t)}_{\lambda_t}$, where the dot denotes the time derivative.
Integrating the above equation and
simplifying it
by noting that $\inpp{\phi_a(t)}{K(t)
\phi_a(t)}_{\lambda_t}=[\inpp{\dot{\phi}_a(t)}
{\phi_a(t)}_{\lambda_t}+\inpp{\phi_a(t)}
{\dot{\phi}_a(t)}_{\lambda_t}]/2$, we obtain
$\alpha=\beta+\gamma$, with
\begin{eqnarray}
\beta &:=& -\int_0^\tau\ud t\inpp{\phi_a(t)}{H(t)
\phi_a(t)}_{\lambda_t},\label{eq:DP}\\
\gamma &:=& -\Im\int_0^\tau\ud t
\inpp{\phi_a(t)}{\dot{\phi}_a(t)}_{\lambda_t}.\label{eq:GP}
\end{eqnarray}
Equation (\ref{eq:DP}) indicates that the phase $\beta$ depends explicitly on the Hamiltonian and thus represents a DP. On the contrary, the phase $\gamma$, as a factor obtained by removing the DP from the total phase, depends solely upon the closed curve $C$ in Eq.~(\ref{eq:curve}), as will be proved shortly. Therefore, $\gamma$ is our GP.

To prove that $\gamma$ is uniquely determined by $C$, we resort to the following gauge-invariant formula of $\gamma$:
\begin{eqnarray}\label{formula:GP}
\gamma=\arg\inpp{\phi(0)}{\phi(\tau)}_{\lambda_0}-
\Im\int_0^\tau\ud t\inpp{\phi(t)}{\dot{\phi}(t)}_{\lambda_t}.\nonumber\\
\end{eqnarray}
Here, $\ket{\phi(t)}\in\mathcal{H}(\lambda_t)$ stands for a gauge which satisfies $\ket{\phi(t)}=e^{i\vartheta(t)}\ket{\phi_a(t)}$ for some real function $\vartheta(t)$. It is easy to verify that $\gamma$ in
Eq.~(\ref{formula:GP}) is independent of specific choices of $\ket{\phi(t)}$. Note that there is a special gauge, denoted by $\ket{\phi_b(t)}$, that satisfies
\begin{eqnarray}\label{eq:pt}
\Im\inpp{{\phi_b}(t)}{\dot{{\phi_b}}
(t)}_{\lambda_t}=0.
\end{eqnarray}
From Eqs.~(\ref{formula:GP}) and (\ref{eq:pt}), it follows immediately that
\begin{eqnarray}\label{eq:sim-GP}
\gamma=\arg\inpp{\phi_b(0)}{\phi_b
(\tau)}_{\lambda_0}.
\end{eqnarray}
On the other hand, noting that $\rho(t)=\ket{\phi_b(t)}\bra{\widetilde{\phi}_b(t)}=\ket{\phi_b(t)}\bra{\phi_b(t)}W(\lambda_t)$, we have, after direct calculations,
\begin{eqnarray}\label{eq:rho-dot}
\dot{\rho}(t)\ket{\phi_b(t)}=\ket{\dot{\phi}_b(t)}
-\inpp{\phi_b(t)}{\dot{\phi}_b(t)}_{\lambda_t}\ket{\phi_b(t)}.
\nonumber\\
\end{eqnarray}
Here, the fact $\bra{\phi_b(t)}\dot{W}(\lambda_t)\ket{\phi_b(t)}=-\inpp{\dot{\phi}_b(t)}
{\phi_b(t)}_{\lambda_t}-\inpp{\phi_b(t)}{\dot{\phi}_b(t)}_{\lambda_t}$ has been used. Rewriting Eq.~(\ref{eq:rho-dot}) by using Eq.~(\ref{eq:pt}) gives
\begin{eqnarray}\label{eq:diff}
\ket{\dot{\phi}_b(t)}=\left[\dot{\rho}(t)+\textrm{Re}
\inpp{{\phi_b}(t)}{\dot{\phi}_b(t)}_{\lambda_t}\right]
\ket{{\phi_b}(t)}.
\end{eqnarray}
Integrating this differential equation, we have
\begin{eqnarray}\label{eq:special-gauge}
\ket{{\phi_b}(t)}=
Te^{\int_0^t\ud s\dot{\rho}(s)}\ket{{\phi_b}(0)}
e^{\int_0^t\ud s
\textrm{Re}
\inpp{{\phi_b}(s)}{\dot{\phi}_b(s)}_{\lambda_s}},\nonumber\\
\end{eqnarray}
where $T$ denotes the time-ordering operator. Substituting Eq. (\ref{eq:special-gauge}) into Eq. (\ref{eq:sim-GP}), we arrive at an expression for $\gamma$ exclusively in terms of $\rho(s)$:
\begin{eqnarray}\label{eq:rho-GP}
\gamma=\arg\tr\left[\rho(0)Te^{\int_0^t\ud s\dot{\rho}(s)}\right].
\end{eqnarray}
Here, the term $e^{\int_0^t\ud s
\textrm{Re}
\inpp{{\phi_b}(s)}{\dot{\phi}_b(s)}_{\lambda_s}}$ has been neglected, since it is a positive number and thus makes no contribution.
Equation (\ref{eq:rho-GP}) clearly shows that $\gamma$ is uniquely determined by $C$ in Eq. (\ref{eq:curve}), thus completing the proof.

It is worth noting that in Ref. \cite{2018Zhang}, we have obtained the GP $\gamma$ for the eigenstates of $H(\lambda)$, i.e., the state appearing in Eq.~(\ref{eq:GP}) is one of these eigenstates.
The GP $\gamma$ obtained in Ref. \cite{2018Zhang} may be regarded as a counterpart of Berry's phase \cite{1984Berry45}. In contrast, the GP $\gamma$ obtained here is for a generic cyclic state, and therefore, it may be regarded as a counterpart of Aharonov-Anandan's phase \cite{1987Aharonov1593}.
Besides, we deduce from Eq. (\ref{eq:rho-GP}) that $\gamma$ is not only gauge-invariant but also reparametrization-invariant, i.e., $\gamma$ is invariant under transformations $s\mapsto s^\prime :=s^\prime(s)$. This implies that our GP is independent of the rate of change of the evolution.

\section{Differential geometry concepts}\label{sec:concepts}

To prepare for the formulations of differential geometry concepts, we introduce several notions needed and then set the stage of our analysis. One of the notions needed is the space of rays for time-dependent $\mathcal{PT}$QM. The subset of the space of rays, denoted as $\mathcal{R}(\lambda)$, is defined to be associated to the physical Hilbert space $\mathcal{H}(\lambda)$. A ray in $\mathcal{R}(\lambda)$, represented by the symbol $[\ket{\psi}]$, is an equivalence class, $[\ket{\psi}]:=\{\ket{\phi}\in\mathcal{H}(\lambda)~|~
\ket{\phi}=c\ket{\psi}~\textrm{for some non-zero $c\in\mathbb{C}$}\}$, obtained by identifying states $\ket{\phi}$ in $\mathcal{H}(\lambda)$ which differ from $\ket{\psi}$ only by an overall rescaling. The space of rays itself, denoted by $\mathcal{R}$, is defined to be the disjoint union of $\mathcal{R}(\lambda)$ over $\lambda\in M$, i.e., $\mathcal{R}:=\bigsqcup_{\lambda\in M}\mathcal{R}(\lambda)$.

There is a one-one correspondence between rays in $\mathcal{R}(\lambda)$ and density operators over $\mathcal{H}(\lambda)$. Indeed, given a ray $[\ket{\psi}]$ in $\mathcal{R}(\lambda)$, one can assign to it a unique density operator $\rho$ over $\mathcal{H}(\lambda)$, which is $\rho=
\ket{\psi}\bra{\widetilde{\psi}}/\inpp{\psi}{\psi}_\lambda$. Conversely, given a density operator $\rho$ over $\mathcal{H}(\lambda)$, one can express it as $\rho=\ket{\psi}\bra{\widetilde{\psi}}$ for some $\ket{\psi}\in\mathcal{N}(\lambda)$. Then, the unique ray associated to $\rho$ is $[\ket{\psi}]$. Under the effect of this one-one correspondence, a curve in $\mathcal{R}$ can be simply understood as a curve of density operators.
Now, it becomes clear that
$C$ in Eq.~(\ref{eq:curve}) is actually a curve in the space of rays $\mathcal{R}$.

To obtain a fiber-bundle structure, we define a projection map, denoted as $\Pi$, which maps a state
$\ket{\psi}\in\mathcal{N}(\lambda)$ to the ray $[\ket{\psi}]\in\mathcal{R}(\lambda)$, i.e., $\Pi(\ket{\psi})=[\ket{\psi}]$. This is a map from the disjoint union of $\mathcal{N}(\lambda)$ over $\lambda\in M$, i.e., $\bigsqcup_{\lambda\in M}\mathcal{N}(\lambda)$, to the space of rays $\mathcal{R}$.
Under the influence of $\Pi$, the disjoint union of $\mathcal{N}(\lambda)$ over $\lambda\in M$ becomes a principle $U(1)$-bundle with $\mathcal{R}$ acting as the base manifold. Indeed, since all states of the kind $e^{i\theta}\ket{\psi}\in\mathcal{N}(\lambda)$, $\theta$ real, are mapped via $\Pi$ to the same ray $[\ket{\psi}]\in\mathcal{R}(\lambda)$,
the fibres on top of a point of $\mathcal{R}$ constitute a $U(1)$ group. Hereafter, we denote by $P(\mathcal{R},U(1))$ this principle $U(1)$-bundle.

In the following, we focus on a local patch on $\mathcal{R}$
and the region of $P(\mathcal{R},U(1))$ over the patch.
Let $(\lambda^1,\cdots,\lambda^m,\lambda^{m+1},\cdots,\lambda
^{m+n})$, where $\lambda^{\mu}\in\mathbb{R}$, $\mu=1,\cdots,m+n$, be the local coordinates of a point of $\mathcal{R}$. Here, $\lambda^{\mu}$, $\mu=1,\cdots,m$, are the system's parameters as before, used to specify which subset, i.e., $\mathcal{R}(\lambda)$, the point belongs to. The remainder $\lambda^{\mu}$, $\mu=m+1,\cdots,m+n$, are used to represent local coordinates parameterizing the manifold $\mathcal{R}(\lambda)$ [$n=\textrm{dim}(\mathcal{R}(\lambda))$]. Then, the local coordinates of a point of $P(\mathcal{R},U(1))$ can be expressed as $(\theta,\lambda^1,\cdots,\lambda^m,\lambda^{m+1},\cdots,\lambda
^{m+n})$, where $\theta\in\mathbb{R}$ is defined up to an integer multiple of $2\pi$ \cite{1note_QGT_Accompanying}.
Using these local coordinates, we can express density operators and states in a coordinate-dependent form. Since a density operator $\rho$ is a point of $\mathcal{R}$, it can be represented as $\rho=\rho(\lambda^1,\cdots,\lambda^{m+n})$. Likewise,
a state $\ket{\phi}$, as a point of $P(\mathcal{R},U(1))$, can be written as $\ket{\phi}=\ket{\phi(\theta,\lambda^1,\cdots,\lambda^{m+n})}$.

After the preparation, we now begin to formulate differential geometry concepts, including connection, curvature, parallel transport, metric tensor, and quantum geometric tensor.

\subsection{Connection}

Consider curves $\ket{\phi(t)}:=\ket{\phi(\theta_t,\lambda_t^1,\cdots,
\lambda_t^{m+n})}$ in $P(\mathcal{R},U(1))$. Associated with each curve $\ket{\phi(t)}$, there is a tangent vector $\ud\ket{{\phi}(t)}/\ud t$. Conversely, every tangent vector can be produced in this way, i.e., by acting the operator $\ud/\ud t$ on a curve $\ket{\phi(t)}$ in $P(\mathcal{R},U(1))$.
A connection on $P(\mathcal{R},U(1))$ is specified by splitting all tangent vectors into vertical and horizontal parts, that is,
$\ket{\dot{\phi}(t)}=\ket{\dot{\phi}_v(t)}+\ket{\dot{\phi}_h(t)}$
\cite{1990Nakahara}.
Like tangent vectors, horizontal parts $\ket{\dot{\phi}_h(t)}$ are produced by acting the covariant derivative $D/\ud t$ on curves $\ket{\phi(t)}$, i.e., $\ket{\dot{\phi}_h(t)}=D\ket{\phi(t)}/\ud t$. The action of the covariant derivative reads $D\ket{\phi(t)}/\ud t=\dot{\lambda}^\mu D_\mu\ket{\phi(t)}$, with $D_\mu:=\partial_{\mu}-A_\mu\partial_\theta$. Here, $\partial_\mu:=\partial/\partial\lambda^\mu$, $\partial_\theta:=\partial/\partial\theta$, and
$A_\mu$ is a real coefficient representing the component of a connection one-form $A:=A_\mu \ud \lambda^\mu$, to be specified later on. Throughout, the Einstein summation convention is assumed, and the summation is understood to be taken over all the indices $\mu=1,\cdots,m+n$.

%

To specify the connection $A$, we adopt the following definition:
\begin{eqnarray}\label{eq:decom}
\ket{\dot{\phi}_h(t)}:=\ket{\dot{\phi}(t)}-
i\Im\inpp{\phi(t)}{\dot{\phi}(t)}_{\lambda_t}
\ket{\phi(t)}.
\end{eqnarray}
Equation (\ref{eq:decom}) implies that the horizontal part satisfies
\begin{eqnarray}\label{para-trans}
\Im\inpp{\phi(t)}{\dot{\phi}_h(t)}_{\lambda_t}=0.
\end{eqnarray}
Inserting $\ket{\dot{\phi}_h(t)}=\dot{\lambda}^\mu D_\mu\ket{\phi(t)}$ into Eq. (\ref{para-trans}) and noting that $\dot{\lambda}^\mu$ can take arbitrary values, we have
$\Im\inpp{\phi(t)}{D_\mu\phi(t)}_{\lambda_t}=0$.
Rewriting this equation by using $D_\mu=\partial_{\mu}-A_\mu\partial_\theta$ gives
\begin{eqnarray}
A_\mu=\Im\inpp{\phi}{\partial_{\mu}\phi}
_{\lambda}/
{\Im\inpp{\phi}{\partial_\theta\phi}_{\lambda}}.
\label{eq:step-A}
\end{eqnarray}
Here, $t$ is omitted for ease of notation. On the other hand, by comparing an infinitesimal $U(1)$ action
$e^{i\delta\theta}\ket{\phi(\theta,\lambda^1,\cdots,\lambda^{m+n})}
=\ket{\phi(\theta,\lambda^1,\cdots,\lambda^{m+n})}
+i\delta\theta\ket{\phi(\theta,\lambda^1,\cdots,\lambda^{m+n})}$
with a Taylor-series expansion $\ket{\phi(\theta+\delta\theta,\lambda^1,\cdots,\lambda^{m+n})}
=\ket{\phi(\theta,\lambda^1,\cdots,\lambda^{m+n})}+
\delta\theta\partial_
\theta\ket{\phi(\theta,\lambda^1,\cdots,\lambda^{m+n})}$, we deduce that $\partial_\theta\ket{\phi(\theta,\lambda^1,\cdots,\lambda^{m+n})}=
i\ket{\phi(\theta,\lambda^1,\cdots,\lambda^{m+n})}$, where we have used the fact $e^{i\delta\theta}\ket{\phi(\theta,\lambda^1,\cdots,\lambda^{m+n})}=
\ket{\phi(\theta+\delta\theta,\lambda^1,\cdots,\lambda^{m+n})}$.
Substituting the above equality, i.e., $\partial_\theta\ket{\phi}=i\ket{\phi}$, into Eq. (\ref{eq:step-A}),
we arrive at the explicit expression of $A_\mu$:
\begin{eqnarray}\label{eq:conection}
A_\mu=\Im\inpp{\phi}{\partial_{\mu}\phi}
_{\lambda},
\end{eqnarray}
defining the connection $A$ with the one-form
$A=A_\mu\ud \lambda^\mu$. The specific form of $A$ computed from Eq.~(\ref{eq:conection}) is gauge-dependent, i.e., depends on the specific choice of $\ket{\phi}$. Indeed, under gauge transformations $\ket{\phi}\rightarrow e^{i\vartheta}
\ket{\phi}$, where $\vartheta=\vartheta(\lambda^1,\cdots,\lambda^{m+n})$, $A$ transforms as $A\rightarrow A+\ud \vartheta$, i.e., as proper gauge potentials.

Equation (\ref{eq:conection}) represents the connection responsible for the appearance of our GP. To see this, we examine the curve $\ket{\phi_a(t)}$, for which $A_\mu=\Im\inpp{\phi_a(t)}{\partial_{\mu}\phi_a(t)}
_{\lambda_t}$. Rewriting Eq. (\ref{eq:GP}) by using $\ud/\ud t=\dot{\theta}\partial_\theta+\dot{\lambda}^\mu\partial_{\mu}$ and $\partial_\theta\ket{\phi_a(t)}=
i\ket{\phi_a(t)}$ gives
\begin{eqnarray}\label{eq:step-G1}
\gamma=\int_0^\tau-\ud\theta -\Im
\inpp{\phi_a}{\partial_{\mu}\phi_a}_{\lambda}\ud  \lambda^\mu.
\end{eqnarray}
Here, $t$ is omitted again for ease of notation.
Since $\ket{\phi_a(\tau)}=\ket{\phi_a(0)}$, the coordinates of $\ket{\phi_a(t)}$ satisfy
$\theta_\tau=\theta_0$ (mod $2\pi$) and $\lambda_\tau^\mu=\lambda_0^\mu$, $\mu=1,\cdots,m+n$. Besides, the relation between $\rho(t)$ in the curve $C$ and $\ket{\phi_a(t)}$, i.e., $\rho(t)=\ket{\phi_a(t)}\bra{\widetilde{\phi}_a(t)}$, implies that the coordinates of $\rho(t)$ are the $\lambda^\mu$-components of coordinates of $\ket{\phi_a(t)}$. That is, the coordinates of $\rho(t)$ in the curve $C$ are $(\lambda_t^1,\cdots,\lambda_t^{m+n})$.
With the above knowledge, we deduce from Eq.~(\ref{eq:step-G1}) that
\begin{eqnarray}\label{fm:GI-con}
\gamma=-\oint_C A ~~~~(\textrm{mod}~~2\pi),
\end{eqnarray}
where $A=A_\mu\ud \lambda^\mu$, with $A_\mu=\Im
\inpp{\phi_a}{\partial_{\mu}\phi_a}_{\lambda}$. Equation (\ref{fm:GI-con}) clearly shows that $A$ in Eq.~(\ref{eq:conection}) is indeed the connection associated with our GP. So, $A$ can be regarded as a counterpart of the Berry connection \cite{1984Berry45}.

\subsection{Curvature}

Using Stokes' theorem, we deduce from Eq. (\ref{fm:GI-con}) that
\begin{eqnarray}\label{fm:GI-cur}
\gamma=-\int_S \Omega~~~~(\textrm{mod}~~2\pi),
\end{eqnarray}
where $S$ is any surface enclosed by the curve $C$, and $\Omega:=\ud A$ represents the curvature associated to our GP, with $\ud$ denoting the exterior derivative. $\Omega$ is
a counterpart of the Berry curvature \cite{1984Berry45}. To obtain an explicit expression of $\Omega$, we insert $A=A_\mu\ud\lambda^\mu$ into $\Omega=\ud A$, and obtain
$\Omega=\partial_\mu A_\nu\ud\lambda^\mu\wedge\ud\lambda^\nu$,
where $\wedge$ denotes the wedge product. Using the anti-commutativity of the wedge product, i.e., $\ud\lambda^\mu\wedge\ud\lambda^\nu=-\ud\lambda^\nu\wedge\ud\lambda^\mu$,
we can rewrite the above expression as
\begin{eqnarray}
\Omega=\frac{1}{2}(\partial_\mu A_\nu-\partial_\nu A_\mu)\ud\lambda^\mu\wedge\ud\lambda^\nu.
\end{eqnarray}
So, the components of $\Omega$ read
\begin{eqnarray}\label{eq:cur-component}
\Omega_{\mu\nu}=\frac{1}{2}(\partial_\mu A_\nu-\partial_\nu A_\mu),
\end{eqnarray}
i.e., $\Omega=\Omega_{\mu\nu}\ud\lambda^\mu\wedge\ud\lambda^\nu$.
Substituting $A_\mu=\Im
\inpp{\phi_a}{\partial_{\mu}\phi_a}_{\lambda}$ into Eq.~(\ref{eq:cur-component}) and noting that $\inpp{\phi_a}{\partial_{\mu}\phi_a}_{\lambda}=
\inp{\widetilde{\phi}_a}{\partial_\mu\phi_a}$, we obtain
\begin{eqnarray}\label{cur}
\Omega_{\mu\nu}=\frac{1}{2}\Im
\left(\inp{\partial_\mu\widetilde{\phi}_a}{\partial_\nu
\phi_a}+\inp{\partial_\mu\phi_a}{\partial_\nu
\widetilde{\phi}_a}\right).
\end{eqnarray}
Here, we have used the fact $\Im\inp{\partial_\nu\widetilde{\phi}_a}{\partial_\mu\phi_a}
=-\Im\inp{\partial_\mu\phi_a}{\partial_\nu
\widetilde{\phi}_a}$.

It is worth noting that $\Omega_{\mu\nu}$ is gauge-invariant. This point can be verified straightforwardly by plugging $\ket{\phi}=e^{i\vartheta}\ket{\phi_a}$ into Eq.~(\ref{cur}). As an immediate consequence, $\ket{\phi_a}$ appearing in Eq. (\ref{cur}) can be replaced by any gauge $\ket{\phi}$ with $\ket{\phi}=e^{i\vartheta}
\ket{\phi_a}$ for some $\vartheta$, that is
\begin{eqnarray}\label{cp:cur}
\Omega_{\mu\nu}=\frac{1}{2}\Im
\left(\inp{\partial_\mu\widetilde{\phi}}{\partial_\nu
\phi}+\inp{\partial_\mu\phi}{\partial_\nu
\widetilde{\phi}}\right).
\end{eqnarray}
It is also worth noting that like the Berry curvature \cite{1984Berry45}, $\Omega_{\mu\nu}$ is a real anti-symmetric tensor, i.e.,
$\Omega_{\mu\nu}=\Omega_{\mu\nu}^*$ and $\Omega_{\mu\nu}=-\Omega_{\nu\mu}$.

\subsection{Parallel transport}\label{para-tran}

A choice of connection is equivalent to a notion of parallel transport. By definition, a curve
$\ket{\phi(t)}$ is said to be parallel transported along a curve in the base manifold if the vertical part of its tangent vector, i.e., $\ket{\dot{\phi}_v(t)}$, vanishes \cite{1990Nakahara}. Using Eq.~(\ref{eq:decom}) and noting the relation $\ket{\dot{\phi}(t)}=
\ket{\dot{\phi}_v(t)}+\ket{\dot{\phi}_h(t)}$, we have
\begin{eqnarray}
\ket{\dot{\phi}_v(t)}=i\Im\inpp{\phi(t)}{\dot{\phi}(t)}_{\lambda_t}
\ket{\phi(t)}.
\end{eqnarray}
Therefore, $\ket{\dot{\phi}_v(t)}$ vanishes if and only if
\begin{eqnarray}\label{para-tran-con}
\Im\inpp{\phi(t)}{\dot{\phi}(t)}_{\lambda_t}=0,
\end{eqnarray}
representing the parallel transport condition associated with the connection $A$. Equation (\ref{para-tran-con}) is
a counterpart of the Berry-Simon parallel transport condition \cite{1983Simon2167}.
It depicts a parallel way of transporting $\ket{\phi(t)}$ along a curve in $\mathcal{R}$.
Evidently, $\ket{\phi_b(t)}$ fulfills Eq.~(\ref{para-tran-con}) and hence is parallel transported. This transport is along the closed curve $C$ in Eq.~(\ref{eq:curve}), since $\ket{\phi_b(t)}\bra{\widetilde{\phi}_b(t)}=\rho(t)$.
Starting at an initial point $\ket{\phi_b(0)}$, the transport will end at a different point $\ket{\phi_b(\tau)}=e^{i\gamma}\ket{\phi_b(0)}$, as can be easily verified by using Eq.~(\ref{eq:sim-GP}). The difference, known as holonomy, is precisely our GP $\gamma$.

\subsection{Metric tensor}

To obtain a metric tensor, we introduce a formula for the fidelity between two nearby density operators $\rho(\lambda^1,\cdots,\lambda^{m+n})$ and $\rho(\lambda^1+\delta\lambda^1,\cdots,\lambda^{m+n}+
\delta\lambda^{m+n})$. It reads
$F(\rho(\lambda^1,\cdots,\lambda^{m+n}),\rho(\lambda^1+\delta\lambda^1,\cdots,\lambda^{m+n}+
\delta\lambda^{m+n}))
:=\tr\abs{\rho^{1/2}
(\lambda^1,\cdots,\lambda^{m+n})\rho(\lambda^1+\delta\lambda^1,\cdots,\lambda^{m+n}+
\delta\lambda^{m+n})
\rho^{1/2}(\lambda^1,\cdots,\lambda^{m+n})}^{1/2}$.
Here, for an operator $X$, $\abs{X}:=\sqrt{X^\prime X}$, with $X^\prime$ being the Hermitian conjugate of $X$ w.r.t. the $\lambda$-dependent inner product. This formula is almost of the same form as that in standard quantum mechanics \cite{2010Nielsen}. Note that $\rho(\lambda^1,\cdots,\lambda^{m+n})=\ket{\phi(\theta,\lambda^1,
\cdots,\lambda^{m+n})}\bra{\widetilde{\phi}(\theta,\lambda^1,
\cdots,\lambda^{m+n})}$ and $\rho(\lambda^1+\delta\lambda^1,\cdots,\lambda^{m+n}+\delta
\lambda^{m+n})=\ket{\phi(\theta,\lambda^1+\delta\lambda^1,
\cdots,\lambda^{m+n}+\delta\lambda^{m+n})}
\bra{\widetilde{\phi}(\theta,\lambda^1+\delta\lambda^1,
\cdots,\lambda^{m+n}+\delta\lambda^{m+n})}$. Here, the coordinate $\theta$ has no effect, since $\ket{\phi(\theta,\lambda^1,
\cdots,\lambda^{m+n})}=e^{i\theta}\ket{\phi(0,\lambda^1,
\cdots,\lambda^{m+n})}$ \cite{2note_QGT_Accompanying}.
Inserting these two expressions into the formula gives
\begin{eqnarray}\label{cp:fidelity}
&&F(\rho(\lambda^1,\cdots,\lambda^{m+n}),\rho(\lambda^1+\delta\lambda^1,\cdots,\lambda^{m+n}+
\delta\lambda^{m+n}))=\nonumber\\
&&\abs{\inp{\widetilde{\phi}(\theta,
\lambda^1+\delta\lambda^1,\cdots,\lambda^{m+n}+\delta\lambda^{m+n})}
{\phi(\theta,\lambda^1,\cdots,\lambda^{m+n})}
\nonumber\\
&&\inp{\widetilde{\phi}(\theta,\lambda^1,\cdots,\lambda^{m+n})}{
\phi(\theta,\lambda^1+\delta\lambda^1,\cdots,\lambda^{m+n}+\delta
\lambda^{m+n})}}^{\frac{1}{2}}.\nonumber\\
\end{eqnarray}

In the spirit of Bures distance \cite{1980Provost289}, the distance element between $\rho(\lambda^1,\cdots,\lambda^{m+n})$ and $\rho(\lambda^1+\delta\lambda^1,\cdots,\lambda^{m+n}+
\delta\lambda^{m+n})$ can be defined as
\begin{eqnarray}\label{cp:metric}
&&\ud s^2:=2[1-\nonumber\\
&&F(\rho(\lambda^1,\cdots,\lambda^{m+n}),\rho(\lambda^1+\delta\lambda^1,\cdots,\lambda^{m+n}+
\delta\lambda^{m+n}))].\nonumber\\
\end{eqnarray}
Substituting Eq.~(\ref{cp:fidelity}) into this defining expression (\ref{cp:metric}) and using Taylor-series expansions of $\ket{\phi(\theta,\lambda^1+\delta\lambda^1,\cdots,\lambda^{m+n}
+\delta\lambda^{m+n})}$ and
$\ket{\widetilde{\phi}(\theta,\lambda^1+\delta\lambda^1,\cdots,\lambda^{m+n}
+\delta\lambda^{m+n})}$, we obtain, up to second order,
\begin{eqnarray}
\ud s^2=g_{\mu\nu}\ud\lambda^{\mu}\ud\lambda^\nu,
\end{eqnarray}
with
\begin{eqnarray}\label{cp:metric-tensor}
g_{\mu\nu}=\frac{1}{2}\Re\left(\inp{\partial_\mu\widetilde{\phi}}{
\partial_\nu\phi}-\inp{\partial_\mu\widetilde{\phi}}{\phi}
\inp{\widetilde{\phi}}{\partial_\nu\phi}
+\textrm{``term $\widetilde{\phi}\leftrightarrow\phi$''}
\right).\nonumber\\
\end{eqnarray}
Here, ``term $\widetilde{\phi}\leftrightarrow\phi$'' stands for $\inp{\partial_\mu{\phi}}{
\partial_\nu\widetilde{\phi}}-
\inp{\partial_\mu{\phi}}{\widetilde{\phi}}
\inp{{\phi}}{\partial_\nu\widetilde{\phi}}$.
The derivation of Eq.~(\ref{cp:metric-tensor}) is the same as that in Ref. \cite{2018Zhang} (See Supplemental Material of Ref. \cite{2018Zhang}). Hence, it is omitted here.
Equation (\ref{cp:metric-tensor}) is the desired metric tensor, which generalizes the one proposed in Ref. \cite{2018Zhang}. Like the seminal one  \cite{1980Provost289} as well as the one proposed in Ref. \cite{2018Zhang}, the metric tensor (\ref{cp:metric-tensor}) is a real symmetric tensor, i.e., $g_{\mu\nu}=g_{\mu\nu}^*$
and $g_{\mu\nu}=g_{\nu\mu}$. Its physical relevance has been shown in Ref. \cite{2018Zhang}.

Our metric tensor reduces to the seminal one \cite{1980Provost289} when $W(\lambda)=I$, where $I$ denotes the identity operator. Hence, $\ud s^2$ may be Riemannian, as the seminal metric is Riemannian. However, as noted in Ref. \cite{2018Zhang}, $\ud s^2$ may also be pseudo-Riemanniana, a case which is absent in standard quantum mechanics. To comprehend this case, one may recall the Minkowski metric $\ud s_M^2$ in special relativity. $\ud s_M^2$ is a pseudo-Riemannian metric describing the geometry of spacetime, according to which a curve of events in spacetime is said to be spacelike, lightlike, or timelike if $\ud s_M^2>0$, $\ud s_M^2=0$, or $\ud s_M^2<0$, respectively. Similar to the Minkowski metric, our metric $\ud s^2$, when being pseudo-Riemannian, results in three types of evolutions, classified according to the sign of $\ud s^2$, i.e., $\ud s^2>0$, $\ud s^2=0$, and $\ud s^2<0$. Resorting to the language of special relativity, we refer to them as spacelike, lightlike, and timelike evolutions, respectively. An example of showing the existence of timelike evolutions will be given later on.

\subsection{Quantum geometric tensor}

In standard quantum mechanics, the seminal QGT \cite{1980Provost289} is a gauge-invariant complex Hermitian tensor, with its imaginary part determining the Berry curvature and its real part inducing a metric tensor on the space of rays. In this paper, we advocate a QGT for time-dependent $\mathcal{PT}$QM.

The QGT advocated here reads
\begin{eqnarray}\label{cp:QGT}
Q_{\mu\nu}=\frac{1}{2}\left(\inp{\partial_\mu\widetilde{\phi}}{
\partial_\nu\phi}-\inp{\partial_\mu\widetilde{\phi}}{\phi}
\inp{\widetilde{\phi}}{\partial_\nu\phi}
+\textrm{``term $\widetilde{\phi}\leftrightarrow\phi$''}
\right).\nonumber\\
\end{eqnarray}
$Q_{\mu\nu}$ in Eq.~(\ref{cp:QGT}) shares all the features of the seminal QGT. First, $Q_{\mu\nu}$ is a gauge-invariant complex Hermitian tensor. Indeed, direct calculations show that
\begin{eqnarray}
Q_{\mu\nu}=&&\frac{1}{2}\tr\left[{\partial_{\mu}(\ket{\phi}
\bra{\widetilde{\phi}})(1-\ket{\phi}
\bra{\widetilde{\phi}})\partial_{\nu}(\ket{\phi}
\bra{\widetilde{\phi}})}\right.\nonumber\\
&&\left.{+\partial_{\mu}(\ket{\widetilde{\phi}}
\bra{\phi})(1-\ket{\widetilde{\phi}}
\bra{\phi})\partial_{\nu}(\ket{\widetilde{\phi}}
\bra{\phi})}
\right],
\end{eqnarray}
from which the gauge-invariance of $Q_{\mu\nu}$ follows immediately. Besides, it is easy to see that $Q_{\mu\nu}=Q_{\nu\mu}^*$, implying that $Q_{\mu\nu}$ is a complex Hermitian tensor. Second, the imaginary part of $Q_{\mu\nu}$ determines the curvature in Eq.~(\ref{cp:cur}).
To see this, note that $\inp{\partial_\mu\widetilde{\phi}}{\phi}=-\inp{\partial_\mu\phi}{
\widetilde{\phi}}^*$ and $\inp{\widetilde{\phi}}{\partial_\nu\phi}=-\inp{\phi}{\partial_\nu
\widetilde{\phi}}^*$. It implies that $-\inp{\partial_\mu\widetilde{\phi}}{\phi}
\inp{\widetilde{\phi}}{\partial_\nu\phi}-
\inp{\partial_\mu\phi}{
\widetilde{\phi}}\inp{\phi}{\partial_\nu
\widetilde{\phi}}$, i.e., a term appearing in Eq.~(\ref{cp:QGT}), is real, which leads to $\Im Q_{\mu\nu}=
\Im(\inp{\partial_\mu\widetilde{\phi}}{\partial_\nu
\phi}+\inp{\partial_\mu\phi}{\partial_\nu
\widetilde{\phi}})/2$. That is,
\begin{eqnarray}\label{QGT-cur}
\Im \left[Q_{\mu\nu}\right]=\Omega_{\mu\nu}.
\end{eqnarray}
Third, the real part of $Q_{\mu\nu}$ induces the metric tensor $g_{\mu\nu}$ in Eq.~(\ref{cp:metric-tensor}), i.e.,
\begin{eqnarray}\label{QGT-metric}
\Re \left[Q_{\mu\nu}\right]=g_{\mu\nu}.
\end{eqnarray}
This point can be easily verified by comparing Eq.~(\ref{cp:metric-tensor}) with Eq.~(\ref{cp:QGT}).

From Eqs. (\ref{QGT-cur}) and (\ref{QGT-metric}), it follows
immediately that the QGT (\ref{cp:QGT}) depicts a unified picture: Its imaginary part gives the Berry curvature (\ref{cp:cur}) and thus further determines the GP (\ref{fm:GI-cur}), whereas its real part induces the metric tensor (\ref{cp:metric-tensor}) and thereby further determines the fidelity (\ref{cp:fidelity}).

\section{On the unconventional geometric phase}\label{sec:unconventional GP}

So far, we have presented our main findings, consisting of a GP and a series of differential geometry concepts, namely, connection, curvature, parallel transport, metric tensor, and QGT.
To exemplify the application of our findings, we revisit one well-known example that yields an interesting GP, called the unconventional GP in the literature~\cite{2003Zhu187902}.

The physical model studied in Ref. \cite{2003Zhu187902} is a harmonic oscillator. Its Hamiltonian reads
\begin{eqnarray}\label{H-UGP}
H(t)=i\Omega_D\left(a^\dagger e^{-i\delta t+i\phi_L}-ae^{i\delta t-i\phi_L}\right),
\end{eqnarray}
where $\Omega_D$, $\delta$, $\phi_L$ are real numbers, and $a^\dagger$ and $a$ are the raising and lowering operators, respectively. The evolving state $\ket{\varphi(t)}$ of the system was shown to be
$\ket{\varphi(t)}=e^{i\gamma(t)}\ket{z(t)}$, provided that the initial state is $\ket{\varphi(0)}=\ket{0}$. Here, $z(t)=i\Omega_D(e^{-i\delta t}-1)e^{i\phi_L}/\delta$,
$\gamma(t)=-\frac{i}{2}\int_0^t\ud s[z^*(s)\dot{z}(s)-\dot{z}^*(s)z(s)]$, and $\ket{z}$ denotes a coherent state. At the time $t=\tau:=2\pi/\delta$, the evolving state $\ket{\varphi(t)}$ returns to its initial physical state, i.e., $\ket{\varphi(\tau)}=e^{i\gamma(\tau)}\ket{0}$, and it acquires a total phase $\gamma(\tau)$. A remarkable observation made in Ref. \cite{2003Zhu187902} is that $\gamma(\tau)$ has a nonzero DP component but is still of geometric nature, i.e., it is an unconventional GP. In showing this, the DP and GP components of $\gamma(\tau)$, denoted respectively by $\gamma_d$ and $\gamma_g$, were calculated, and found to satisfy
$\gamma_d=\eta\gamma_g$ $(\eta\neq 0,-1)$.
So, $\gamma(\tau)=(1+\eta)\gamma_g$, indicating that $\gamma(\tau)$ is of geometric nature as it inherits geometric features from $\gamma_g$. Despite this interesting observation, it remains an open question whether the unconventional GP itself admits a geometric interpretation or not.

To answer this question, we resort to the equivalence of a $\mathcal{PT}$-symmetric system with exact $\mathcal{PT}$-symmetry and a Hermitian system \cite{2013Gong485302}.
Consider the $\mathcal{PT}$-symmetric system with $H(z^1)=0$ and $W(z^1)=e^{2z^{1*}a}e^{2{z^1}a^\dagger}$. Here, the manifold $M$ is the complex plane and its point is designated by $z^1$. An evolution of the system is induced by a curve $z^1=z^1(t)$ and determined purely by the gauge field $K(t)$, that is,
\begin{eqnarray}
i\partial_t\ket{\psi(t)}=iK(t)\ket{\psi(t)},
\end{eqnarray}
with
\begin{eqnarray}
K(t)=-\left[\dot{z}^1(t)a^\dagger+\dot{z}^{1*}(t)a+
2z^1(t)\dot{z}^{1*}(t)\right].
\end{eqnarray}
Under the map
\begin{eqnarray}
\ket{\psi(t)}\rightarrow
\ket{\varphi(t)}:=e^{2z^1(t)a^\dagger}\ket{\psi(t)},
\end{eqnarray}
the $\mathcal{PT}$-symmetric system transforms into an equivalent Hermitian system, i.e.,
\begin{eqnarray}
i\partial_t\ket{\varphi(t)}=h(t)\ket{\varphi(t)},
\end{eqnarray}
with the Hamiltonian
\begin{eqnarray}\label{PTH-UGP}
h(t)=i\left[\dot{z}^1(t)a^\dagger-\dot{z}^{1*}(t)a\right].
\end{eqnarray}
Physically speaking, the $\mathcal{PT}$-symmetric system and its equivalent Hermitian system may be considered two different interpretations of the same physical system.

Suppose now that $z^1(t)=i\Omega_D(e^{-i\delta t}-1)e^{i\phi_L}/\delta$, i.e., $z^1(t)=z(t)$. For this, Eq. (\ref{PTH-UGP}) reduces to Eq. (\ref{H-UGP}). Hence, the evolution of the Hermitian system is simply the evolution process studied in Ref. \cite{2003Zhu187902}. As another interpretation of the same physical system, the $\mathcal{PT}$-symmetric system undergos the corresponding  evolution $\ket{\psi(t)}=e^{-2z^1(t)a^\dagger}\ket{\varphi(t)}$. Since $z^1(\tau)=z^1(0)=0$ and $\ket{\varphi(\tau)}=e^{i\gamma(\tau)}\ket{\varphi(0)}$, this evolution is cyclic, and the evolving state $\ket{\psi(t)}$ of the $\mathcal{PT}$-symmetric system
acquires the same total phase $\gamma(\tau)$ as that of the Hermitian system. Note that for the $\mathcal{PT}$-symmetric system, the total phase accumulated in any cyclic evolution is simply the GP $\gamma$ in Eq. (\ref{eq:GP}), due to the vanishing of its Hamiltonian. Hence, $\gamma(\tau)=\gamma$, i.e., the unconventional GP is precisely the GP expressed by Eq. (\ref{eq:GP}).

To shed more light on the unconventional GP, we calculate the QGT in Eq. (\ref{cp:QGT}), with which, we further obtain the curvature $\Omega$ and the metric $\ud s^2$. To do this, we find the evolution operator of the $\mathcal{PT}$-symmetric system.  Using magnus expansion \cite{2009Blanes151} and noting that the commutator of $K(t)$ at different time is a number, we have that the evolution operator reads $e^{-2z^1(t)a^\dagger}D(z^1(t))$, up to a global phase factor, where $D(z^1):=e^{z^1a^\dagger-z^{1*}a}$ is the displacement operator. So, starting at an arbitrary coherent state $\ket{\textrm{``some complex number''}}$, the evolving state $\ket{\psi(t)}$ reads $\ket{\psi(t)}=e^{-2z^1(t)a^\dagger}\ket{z_1(t)+\textrm{``some complex number''}}$, up to a phase factor. So, the evolving state is of the form $e^{-2z^1a^\dagger}\ket{z^2}$, where $z^2=z^1(t)+\textrm{``some complex number''}$. Substituting $e^{-2z^1a^\dagger}\ket{z^2}$ into Eq.~(\ref{cp:QGT}), i.e.,
setting $\ket{\phi}$ and $\ket{\widetilde{\phi}}$ appearing in Eq.~(\ref{cp:QGT}) as $\ket{\phi}=e^{-2z^1a^\dagger}\ket{z^2}$ and $\ket{\widetilde{\phi}}=e^{2z^{1*}a}\ket{z^2}$, we obtain, after tedious but straightforward calculations,
\begin{eqnarray}
(Q_{\mu\nu})=
\begin{pmatrix}
  0 & 0 & -1 & -i \\
  0 & 0 & i & -1 \\
  -1 & -i & 1 & i \\
  i & -1 & -i & 1
\end{pmatrix}.
\end{eqnarray}
Here, the real coordinates are $(\lambda^1,\lambda^2,\lambda^3,\lambda^4)$ such that $z^1=\lambda^1+i\lambda^2$ and $z^2=\lambda^3+i\lambda^4$.
Now, using Eq. (\ref{QGT-cur}), we easily obtain
\begin{eqnarray}
(\Omega_{\mu\nu})=
\begin{pmatrix}
  0 & 0 & 0 & -1 \\
  0 & 0 & 1 & 0 \\
  0 & -1 & 0 & 1 \\
  1 & 0 & -1 & 0
\end{pmatrix}.
\end{eqnarray}
That is,
\begin{eqnarray}\label{ex:Omega}
\Omega=-2\ud \lambda^1\wedge\ud \lambda^4+2\ud \lambda^2\wedge\ud \lambda^3+2\ud \lambda^3\wedge\ud \lambda^4.
\end{eqnarray}
For the evolution process studied in Ref. \cite{2003Zhu187902}, in which the initial state is $\ket{0}$, we have $z^1(t)=z^2(t)$, leading to the constraints $\lambda^1=\lambda^3$ and $\lambda^2=\lambda^4$. Substituting $\lambda^1=\lambda^3$ and $\lambda^2=\lambda^4$ into Eq.~(\ref{ex:Omega}), we have $\Omega=-2\ud \lambda^1\wedge\ud \lambda^2$. From Eq.~(\ref{QGT-cur}), it follows that
\begin{eqnarray}
\gamma(\tau)=2\iint \ud \lambda^1\wedge\ud \lambda^2,
\end{eqnarray}
representing twice the area enclosed by the curve $z^1(t)$. So, the geometric nature of the unconventional GP is confirmed. Moreover, from Eq.~(\ref{QGT-metric}), we deduce that
\begin{eqnarray}
(g_{\mu\nu})=
\begin{pmatrix}
  0 & 0 & -1 & 0 \\
  0 & 0 & 0 & -1 \\
  -1 & 0 & 1 & 0 \\
  0 & -1 & 0 & 1
\end{pmatrix}.
\end{eqnarray}
That is,
\begin{eqnarray}\label{ex:metric}
\ud s^2=-2\ud \lambda^1\ud \lambda^3-2\ud \lambda^2\ud \lambda^4+\ud \lambda^3\ud \lambda^3+\ud \lambda^4\ud \lambda^4.
\end{eqnarray}
Two of the eigenvalues of $(g_{\mu\nu})$ are positive, i.e., $\frac{1}{2}(1+\sqrt{5})$, whereas the rest are negative, i.e., $\frac{1}{2}(1-\sqrt{5})$. Hence, the metric $\ud s^2$ is pseudo-Riemannian.
Substituting $\lambda^1=\lambda^3$ and $\lambda^2=\lambda^4$ into Eq.~(\ref{ex:metric}),
we have $\ud s^2=-{\ud \lambda^1}{\ud \lambda^1}-{\ud \lambda^2}{\ud \lambda^2}<0$, indicating that the evolution process studied in Ref. \cite{2003Zhu187902} is timelike.

\section{Remarks and conclusion}\label{sec:conclusion}

Before concluding, we make a few brief remarks.
In the accompanying paper \cite{2018Zhang}, we have obtained the same series of geometric concepts as in this paper, but for the eigenstates of $H(\lambda)$. The rather formal treatment in this paper further strengthens the geometric concepts proposed in Ref. \cite{2018Zhang}. Indeed, there is a natural map $f:M\rightarrow \mathcal{R}$, assigning a point $\lambda\in M$ to the density operator $\ket{\Psi_n(\lambda)}\bra{\Phi_n(\lambda)}\in \mathcal{R}$, where $\ket{\Psi_n(\lambda)}$ and $\ket{\Phi_n(\lambda)}$ are the $n$-th right and left eigenstates of $H(\lambda)$, respectively. This map induces a pullback bundle $f^*P(\mathcal{R},U(1))$ \cite{1990Nakahara}, which is a principle $U(1)$-bundle over the base manifold $M$. Thanks to this pullback bundle, all the geometric concepts proposed in Ref. \cite{2018Zhang} can be provided with differential-geometry interpretations, just as those formulated in this paper can, e.g., the Berry curvature in Ref. \cite{2018Zhang} can be interpreted as a local curvature two-form on the pullback bundle.

It is interesting to note that our GP is seemingly similar to, but actually different from, Garrison and Wright's (GW's) GP \cite{1988Garrison177,1990Dattoli5795}.
In the present setting as well as notations, GW's DP and GP can be expressed as
\begin{eqnarray}
\beta &=& -\int_0^\tau\ud t\inpp{\phi_a(t)}{[H(t)+iK(t)]
\phi_a(t)}_{\lambda_t},\label{eq:GW-DP}\\
\gamma &=& -\int_0^\tau\ud t
\inpp{\phi_a(t)}{\dot{\phi}_a(t)}_{\lambda_t},\label{eq:GW-GP}
\end{eqnarray}
respectively. Comparing Eqs. (\ref{eq:DP}) and (\ref{eq:GP}) with Eqs. (\ref{eq:GW-DP}) and (\ref{eq:GW-GP}), one can see that the difference lies in the term $-\int_0^\tau\ud t\inpp{\phi_a(t)}{iK(t)\phi_a(t)}$. That is, Garrison and Wright consider this term as a part of their DP, whereas we treat it as a part of our GP. In this paper, we have shown, from various points of view, that our GP is of geometric nature. For example, it is the integral of a connection one-form (see Eq.~(\ref{fm:GI-con})), the integral of a curvature two-form (see Eq.~(\ref{fm:GI-cur})), and the holonomy of a parallel transport (see the discussion in subsection \ref{para-tran}). Moveover, it has been pointed out that the gauge field $K(t)$ has a geometric origin from a metric-compatible connection of an Hermitian vector bundle \cite{2018Mostafazadeh46022}. This also indicates that the term $-\int_0^\tau\ud t\inpp{\phi_a(t)}{iK(t)\phi_a(t)}$ is of geometric nature. For these reasons, we argue that our definition of GP is more reasonable than GW's GP for the setting under consideration.
Besides, there have been many other formulations of GPs in the literature \cite{1984Wilczek2111,1988Anandan171,
1988Samuel2339,1986Uhlmann229,
2000Sjoeqvist2845,2003Filipp50403,
2003Carollo160402,2004Tong80405,2005Wu140402,2008Maamache150407}.  In contrast to these formulations, the distinct element here is a GP involving a varying Hilbert space $\mathcal{H}(\lambda_t)$ along the path $\lambda_t$.

In conclusion, we have presented a series of results regarding geometric aspects of time-dependent $\mathcal{PT}$QM. Specifically, they are the GP in Eq.~(\ref{eq:GP}), the connection in Eq.~(\ref{eq:conection}), the curvature in Eq.~(\ref{cp:cur}), the parallel transport condition in Eq.~(\ref{para-tran-con}), the metric tensor in Eq.~(\ref{cp:metric-tensor}), and the QGT in Eq.~(\ref{cp:QGT}).
The GP emerges naturally from cyclic evolutions of $\mathcal{PT}$-symmetric systems, and it may be regarded as a counterpart of Aharonov-Anandan's phase. The connection and curvature are responsible for the appearance of the GP, as expressed by Eqs. (\ref{fm:GI-con}) and (\ref{fm:GI-cur}).
The QGT is a unifying concept, of which the imaginary part gives the curvature and the real part induces the metric tensor, as described by Eqs. (\ref{QGT-cur}) and (\ref{QGT-metric}), respectively.

Our results constitute a useful, perhaps indispensible, tool to
tackle physical problems involving $\mathcal{PT}$-symmetric systems with varying system's parameters.
As an illustration of their usefulness, we have solved the open question whether the unconventional GP admits a geometric interpretation or not. Specifically, we have shown that the unconventional GP, instead of being the sum of a DP and a GP, can be expressed as the single GP in Eq. (\ref{eq:GP}), thus making its geometric nature undoubtedly clear.

In passing, we have found for the first time the pseudo-Riemannian feature of the metric tensor in $\mathcal{PT}$-symmetric systems.  As a result, there are now three types of evolutions, i.e.,
spacelike, lightlike, and timelike. The implication of this finding may be an interesting issue for future work.

\begin{acknowledgments}
J.G.~is supported by Singapore Ministry of Education Academic
Research Fund Tier I (WBS No.~R-144-000-353-112) and by the
Singapore NRF grant No. NRFNRFI2017-04 (WBS No. R-144-000-378-281).
Q.W.~is supported by Singapore Ministry of Education Academic
Research Fund Tier I (WBS No.~R-144-000-352-112).
D.-J.~Z.~acknowledges support from the National Natural Science Foundation of
China through Grant No.~11705105 before he joined NUS.
\end{acknowledgments}


\begin{thebibliography}{51}
\bibitem [{1no()}]{1note_QGT_Accompanying}%
  \BibitemOpen
  \href@noop {} {}\bibinfo {note} {Strictly speaking, as a local coordinate,
  $\theta$'s shoud comprise an open interval of $\mathbb{R}$ or $\mathbb{R}$
  itself. This can be achieved by subdividing the local patch into smaller ones
  and then defining local coordinates respectively on these smaller patches. In
  this paper, we do not adopt such a cumbersome treatment and simply define
  $\theta$ up to an integer multiple of $2\pi$.}\BibitemShut {Stop}%
\bibitem [{2no()}]{2note_QGT_Accompanying}%
  \BibitemOpen
  \href@noop {} {}\bibinfo {note} {In mathematical terminology, $\ket{\phi}$
  appearing in the expression $\rho=\ket{\phi}\bra{\widetilde{\phi}}$ can be
  chosen as a local section. A local section is a continuous mapping of a patch
  of $\mathcal{R}$ into the fibers above the patch. This amounts to the fact
  that the local coordinate $\theta$ is a function of
  $(\lambda^1,\cdots,\lambda^{m+n})$, i.e.,
  $\theta=\theta(\lambda^1,\cdots,\lambda^{m+n})$. So, a local section
  $\ket{\phi}$ can be expressed as
  $\ket{\phi}=\ket{\phi(\lambda^1,\cdots,\lambda^{m+n})}$.}\BibitemShut {Stop}%
\end{thebibliography}
%

\end{document}